# Real-time evolution of the buckled Stone-Wales defect in graphene

L.A. Openov [*], A.I. Podlivaev

National Research Nuclear University "MEPhI", Kashirskoe sh. 31, Moscow 115409, Russian Federation

[*] Corresponding author.

E-mail addresses: LAOpenov@mephi.ru, laopenov@gmail.com

# ABSTRACT

Dynamics of the buckled Stone-Wales defect in graphene is studied by means of computer simulation. Thermally activated switching between two degenerate sine-wave-like configurations of the defect is traced in real time. Transition trajectory is found to be rather complex and pass through a multitude of near-planar, wave-like, and irregular configurations. Surprisingly, the switching time fluctuates strongly and can be up to an order of magnitude longer or shorter than the value given by the Arrhenius formula. This is due to a peculiar shape of the potential relief in the neighborhood of sine-wave-like configurations and, as a result, the occurrence of two radically different characteristic times.

# 1. Introduction

Graphene, a hexagonal carbon monolayer [1], is in the focus of the modern condensed matter physics because of both its unique electronic characteristics (Dirack fermions) and potential for practical use [2-4]. As in the case with three-dimensional crystals, the physical properties of graphene are strongly affected by structural imperfections, such as vacancies, substitutional impurities, etc., which are either spontaneously produced at the stage of preparation or generated intentionally (e.g., under particle irradiation). Being two-dimensional, graphene can hold the so called topological defects that are absent in bulk materials. These are intrinsic defects formed upon rearrangement of interatomic bonds. The simplest defect of this kind is the Stone-Wales (SW) defect. It is obtained by in-plane rotating one of the C-C bonds by $90^0$ (SW transformation [5]), so that four hexagons are transformed in a pair of pentagons and a pair of heptagons, see Fig.1. SW defects in graphene have been observed experimentally [6] and predicted to modify its electronic properties and chemical reactivity [7, 8].

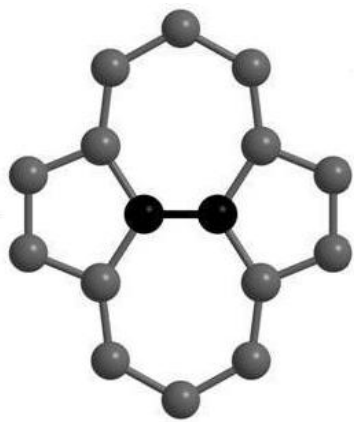

Fig. 1. Top view of the flat SW defect in graphene. The rotated C-C bond is shown in black.

For a long time, SW defects in graphene were thought to be two-dimensional [9, 10] (Fig. 1). However recently Ma et al. [11] have shown with first principles calculations that the flat SW defect is not a local minimum on the potential-energy surface (PES), but a saddle point. The lowest-energy atomic configuration of the SW defect has been shown [11] to have a sine-wave-like form. In this buckled structure, the atoms of the rotated C-C bond (the defect core) move out of plane in the opposite direction, giving rise to the long-range out-of-plane displacement field (Fig. 2). Such buckled SW defects can be responsible (at least partly) for the rippling behavior of graphene [12].

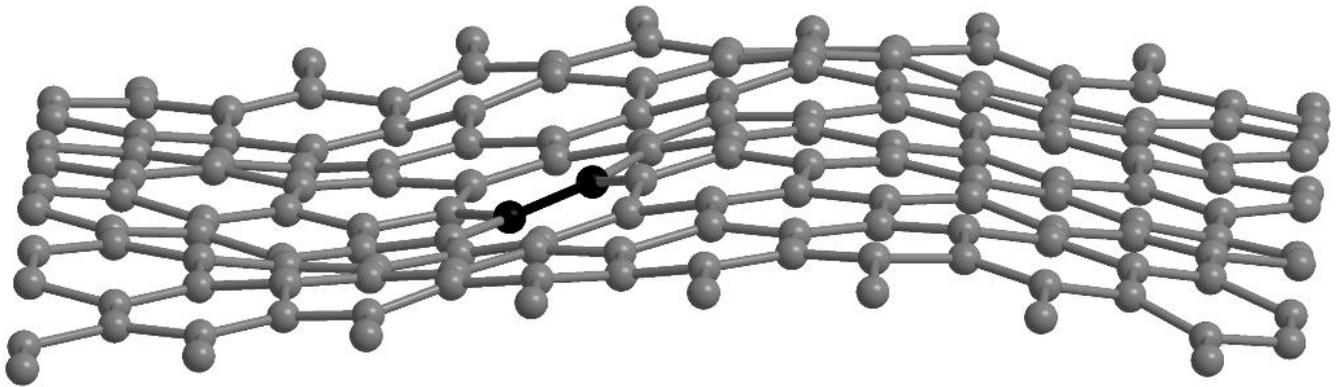

Fig. 2. Perspective view of the sine-wave-like SW defect in graphene. The rotated C-C bond is shown in black.

Since there are two degenerate sine-wave-like configurations (differing in sign of out-of-plane atomic coordinates, i. e., in phase) the switching between them can hinder the observation of the buckled defect structure by means of scanning tunneling microscopy and atomic force microscopy. According to Ref. 11, in order to attain the switching time ~ 1 min sufficient for implementation of the local probe techniques, one should cool the sample down to ~ 70 K. This conclusion rests on the assumption that switching of the SW defect occurs through a cosine-wave-like configuration in which the core atoms are displaced out of plane in the same direction. Note however that in the neighborhood of the sine-wave-like configuration, the PES is rather complex and bears three saddle points with similar energies. One of those saddle points relates to the flat configuration, and two other arise from the cosine-wave-like configurations considered as transition states in Ref. 11. In such a situation, evaluation of the switching time with the Arrhenius formula is questionable.

The main objective of this work is to study the thermally activated evolution of the SW defect by means of molecular dynamics (MD) simulation, to find the transition paths, and to determine the switching times at several selected temperatures. Since the first principles MD is extremely time-consuming, in our calculations we make use of the tight-binding model. The paper is organized as follows. The computational details are briefly described in Sec. 2. The results of MD simulation and PES analysis accompanied by discussion are presented in Sec. 3. We summarize and conclude in Sec. 4.

## 2. Computational details

We modeled graphene by a 160-atom supercell composed of 8 x 5 rectangular 4-atom cells. Periodic boundary conditions along two in-plane directions were employed. To compute the supercell energy at given atomic coordinates, we made use of the non-orthogonal tight-binding potential [13]. Such an approach is much less computer time demanding as compared with first principles methods and provides a possibility to examine the evolution of a system consisting of ~ 100 atoms for a sufficiently long (on the atomic scale) time 1-10 ns [14], the characteristic times for *ab initio* MD being 1-10 ps [15]. For the structure and energetics of carbon clusters (including fullerenes) and bulks (diamond, graphene) this model gives the results in a good agreement with experimental data and first principles calculations [13]. Previously, we successfully used it to study the static and dynamic characteristics of various carbon-based nanostructures [14, 16-18].

MD simulations have been carried out using the velocity Verlet method with the time step 0.272 fs, about two orders of magnitude shorter than the period of the graphene's highest frequency vibrations (see Refs. 14, 16 for details). We recorded the coordinates of all atoms in the supercell every ten MD steps and then visualized the system evolution as a "computer animation". To explore the PES, we employed techniques described in Refs. 19, 20.

### 3. Results and discussion

First, to validate the applicability of our tight-binding model to the structure and energetics of the SW defect in graphene, we calculated several basic characteristics of the defect and compared them with available density functional theory (DFT) data. For the energy barrier to the formation of a flat SW defect we obtained $E_b$ = 8.6 eV, close to the DFT value $E_b$ = 9.2 eV [10]. The formation energy of this defect (the difference in energies of the supercell with and without defect) was found to be $E_{flat}$ = 4.96 eV, in agreement with the DFT result $E_{flat}$ = 5.02 eV [11]. The formation energy of the sine-wave-like SW defect $E_{\text{sine}}$ = 4.51 eV appeared to be little less than the DFT energy $E_{\text{sine}}$ = 4.66 eV computed for the supercell of the same size [11]. For the sine-wave-like configurations, the out-of-plane coordinates of atoms in the rotated C-C bond are $z_{1,2}$ = ± 0.28 Å or ∓ 0.28 Å, while the height between the highest and lowest atom in the supercell is $\Delta z$ = 1.84 Å ($\Delta z$ = 1.51 Å in DFT [11]). In the cosine-wave-like configurations, both $z_1$ and $z_2$ are equal to 0.63 Å or -0.63 Å, and $\Delta z$ = 1.44 Å ($\Delta z \approx$ 1.4 Å in DFT [11]).

Taking the energy of the sine-wave-like configurations as zero, we have for the energy of the cosine-wave-like configurations (first order saddle points of PES) $E_{S1}$ = 0.28 eV, and for the energy of the flat configuration (second order saddle point of PES) $E_{S2}$ = 0.45 eV. These values agree with the DFT results $E_{S1} \approx$ 0.2 eV and $E_{S2} \approx$ 0.4 eV [11]. If one assumes that the switching between sine-wave-like configurations occurs via either of cosine-wave-like configurations as transition states, then the temperature dependence of the switching time $\tau$ is given by the Arrhenius law

$$\frac{1}{\tau \, T} = A \exp\left[-\frac{U}{k_B T}\right], \tag{1}$$

where $k_B$ is the Boltzmann constant, $U = E_{S1}$ is the switching activation energy, and $A$ is the frequency factor. The value of $A$ can be computed from the Vineyard formula [21] whose validity has been explicitly demonstrated in Ref. 22 by the example of several typical thermally activated processes in carbon clusters and nanostructures. In the case under consideration, this formula reads

$$A = 2 \frac{\prod_{i=1}^{3N-3} \nu_i}{\prod_{i=1}^{3N-4} \nu'_i}, \tag{2}$$

where $\nu_i$ are the eigenfrequencies of vibrations of the supercell in the state corresponding to the minimum of the potential energy, $\nu'_i$ are the frequencies of vibrations at the saddle point corresponding to the maximum of the potential energy for one normal coordinate and to the minimum for all the other coordinates (since one of the $3N - 3$ frequencies $\nu'_i$ is imaginary, it is not included in the denominator of Eq.2 for $A$), $N$ is the number of atoms in the supercell, and the factor 2 comes from two equivalent transition pathways. From Eq. 2 we obtain $A = 2.2 \cdot 10^{13}$ s$^{-1}$. In what follows, we neglect the possible dependence of $A$ on temperature.

With the values of $U$ and $A$ at hand, from Eq. 1 one can find the switching time $\tau$ at a given temperature. For example, at $T = 600$ K we have $\tau = 10.2$ ps. Let us compare this value with the MD results. First recall that every sine-wave-like configuration is specified by the signs of transverse coordinates of the core atoms, $z_1$ and $z_2$ (+/- or -\+). That is why we took the switching time equal to the time of the system evolution from the state with $z_1>0$, $z_2<0$ to the state with $z_1<0$, $z_2>0$, and vice versa. The mean value of $\tau$ and the standard error $\Delta \tau$ were obtained by averaging over all switching events observed in a simulation time $t_{sim}$ (usually 0.1-1 ns).

At $T = 600$ K we found $\tau = 16.6 \pm 6.5$ ps (eight switchings, $t_{sim} \sim 100$ ps), in good agreement with Eq. 1. Note, however, that the relative error is rather large ($\approx 40$ %).

The reason lies in the wide scatter of times $\tau_i$ for successive switchings. It is particularly remarkable that the values of $\tau_i$ not only vary greatly but are clustered in two groups corresponding to "fast" and "slow" switchings, $\tau_i \sim 1$ ps and $\sim 30$ ps respectively, four switchings in each group. As a consequence, every switching occurs in a time either much shorter or much longer than that given by Eq. 1, so that the Arrhenius law holds only on the average.

As the temperature increases, the relative number of "slow" switchings decreases along with their times. For example, at $T = 800$ K we detected only one "slow" switching ($\tau_i \sim 15$ ps) per eight "fast" switchings. At $T = 1000$ K there are no "slow" switchings at all, while the times of "fast" switchings decrease moderately. As for the Arrhenius law, it holds (within the statistical error) up to at least $T = 1200$ K. At higher temperatures, the switching times approach the characteristic period of lattice vibrations, and their reliable determination becomes problematic [23].

Now we turn to the case of relatively low temperatures $T = 400\text{-}500$ K for which Eq. 1 gives the switching times $\tau \sim 100$ ps that are still accessible to direct computer simulation. At $T = 500$ K we found one "slow" switching ($\tau = 95$ ps) and three "fast" ones, so that $\tau = 25 \pm 20$ ps. Although this again agrees with the value of $\tau = 32$ ps derived from Eq. 1, the relative error is now almost 100 %. Noteworthy, once the last switching had occurred, we observed no switchings over a time of $\sim 300$ ps, until the simulation was interrupted. Lowering the temperature down to 450 K and next to 400 K resulted in an utter absence of both "slow" and "fast" switchings, even though the simulation time ~ 800 ps far exceeded the switching times expected from Eq. 1, $\tau = 66$ ps at $T = 450$ K and $\tau = 162$ ps at $T = 400$ K.

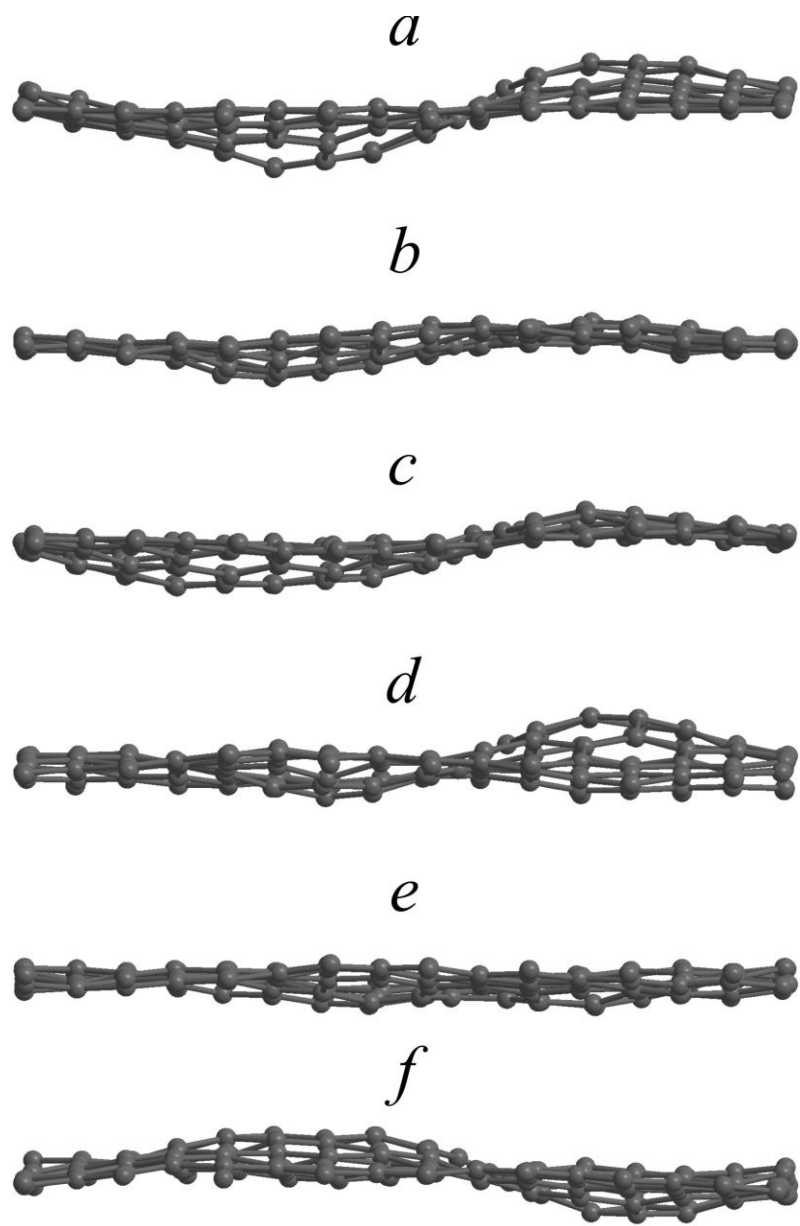

Fig. 3. Several supercell snapshots taken in the course of SW defect evolution from the sine-wave-like configuration formed at a preceding switching to the configuration displaced in phase by half a period. $t = 0$ (*a*), 9 ps (*b*), 74 ps (*c*), 76 ps (*d*), 93 ps (*e*), 95 ps (*f*). Side view. $T = 500$ K.

Trying to get some insight into why the switching time can be more than an order of magnitude longer than that given by the Arrhenius law, we have traced the SW defect evolution from one sine-wave-like configuration to the other during the process of "slow" switching at $T = 500$ K. The results are presented in Fig. 3. One can see that the defect goes through a variety of wave-like, near-planar and irregular configurations. The same picture is observed for "slow" switchings at $T > 500$ K as well as for the switchingless defect evolution at $T = 400$ and 450 K. However, a purely geometrical analysis does not give a clue to understanding the reason for a departure, to say the least, from the Arrhenius law.

Valuable information can be extracted from examination of the PES in the neighborhood of the sine-wave-like configuration. However, the PES is a multidimensional object, Hence, in order to make a consideration more illustrative, it is desirable to restrict ourselves to the analysis of the system energy as a function of as few coordinates as possible. For present purposes, the most relevant coordinates seem to be the transverse coordinates $z_1$, $z_2$ of the core atoms. The switching between the sine-wave-like configurations manifests itself as a change in sign of both $z_1$ and $z_2$ with the constraint $sgn(z_1) = -sgn(z_2)$, see above. Since every defect configuration corresponds to a specific point in the $(z_1, z_2)$ plane, the temperature evolution of the defect can be envisioned as a motion of that point. Fig. 4a shows such "defect projections" recorded at $T = 500$ K each 200 MD steps ($\approx 50$ fs). As is seen, they form two distinct branches corresponding to the prolonged defect evolution before the switching (see Fig.3) and after it. Each branch is oriented parallel to the main diagonal, includes one of two sine-wave-like configurations, and is close to both cosine-wave-like configurations. The switching itself can be viewed as a transition from one branch to another. For the switchingless defect evolution at both $T = 450$ and 400 K, there is just one branch of "projections", see Fig. 4b.

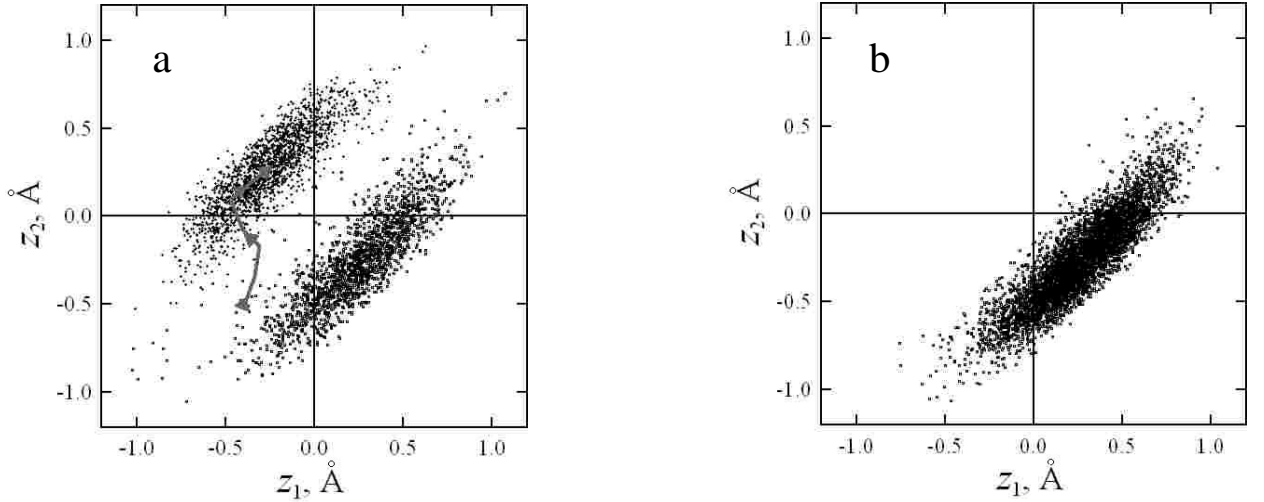

Fig. 4. Illustration of SW defect evolution by projecting the defect configurations onto the ($z_1$, $z_2$) plane, where $z_1$ and $z_2$ are the transverse coordinates of atoms in the rotated C-C bond. They were recorded during the MD simulation at 54.4-fs intervals. The projections corresponding to the sine-wave-like configurations are located in the $z_1 < 0$, $z_2 > 0$ and $z_1 > 0$, $z_2 < 0$ regions (see the text).

(a) $T = 500$ K. Open circles (the lower right) and pluses (the upper left) correspond to configurations emerged upon the defect evolution prior to switching ($t = $ 0-95 ps, see Fig. 3) and after it ($t = $ 95-190 ps) respectively. Note that we observed no switchings at $t = $ 190-370 ps as well. Gray arrows show the evolution path around the instant the defect switches. It corresponds to the time lapse ~ 0.1 ps.

(b) $T = 450$ K. There are no switchings over the course of 816 ps.

The results obtained cause us to suggest the existence of two valleys in the slice of the defect PES along the ($z_1$, $z_2$) plane. Calculations show that this is so indeed, see Fig. 5. Thus the defect evolution at low temperatures (i. e., at low excitation energies) takes place mainly in a severely bounded region of a coordinate space. Besides, the transitions from one valley to another are likely to occur through rather narrow channels containing the saddle points. Taken together, these may be the reason for the apparent departure from the Arrhenius law, most probably because of peculiar temperature dependence of the frequency factor.

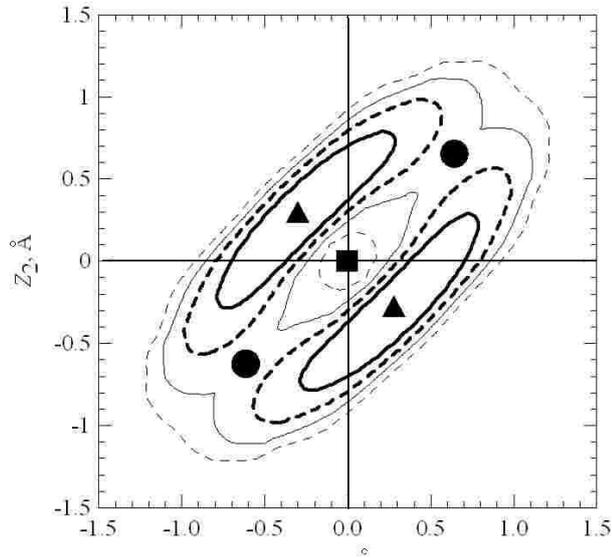

Fig. 5. The constant energy lines for the slice of the SW defect PES along the ($z_1$, $z_2$) plane, where $z_1$ and $z_2$ are out-of-plane coordinates of atoms in the rotated C-C bond. The energy $E_{pot}$ of the sine-wave-like configurations (triangles) is taken for zero. The circles mark the cosine-wave-like configurations (first-order saddle points) with $E_{pot}$ =0.28 eV, and the square stands for the flat configuration (second-order saddle point) with $E_{pot}$ = 0.45 eV. Thick solid, thick dashed, thin solid, and thin dashed lines correspond to $E_{pot}$ = 0.1, 0.2, 0.3, and 0.4 eV respectively.

That the defect switchings are classified as "fast" and slow, may be roughly understood in terms of defect "projection" motion across the ($z_1$, $z_2$) plane as follows. Upon transition to another valley (see Fig. 4a), there are two alternative ways of subsequent defect evolution. On the one hand, as long as, first, the defect still preserves the memory of preceding sine-wave-like configuration and, second, the "projection" remains not far away from the saddle point, the probability of its reverse transition is high, facilitating "fast" defect switching in a time comparable with a period of sine-wave-like configuration oscillations (~ 1 ps). On the other hand, in the case that the defect shape has managed to change noticeably towards another (displaced in phase by half a period) sine-wave-like configuration prior to reverse transition occurred, the "projection" goes on a long journey over the valley until favorable conditions for the next switching arise. As indicated above, the switching time in this case far exceeds that given by the Arrhenius law. So, there are two characteristic switching times, one of which (short) weakly depends on temperature and the other (long) grows steeply as temperature decreases. At low temperatures, those times differ by orders of magnitude.

Finally, we briefly comment on the switching criterion used in this work (the change of signs of $z_1$ and $z_2$ coordinates from +/- to -\+ or vice versa). It can be argued that such a change is not by itself an indication of switching to another sine-wave-like configuration since one should in addition wait until an appropriate overall defect shape is formed. This is in part true. However, we have checked that relaxation of the supercell with $\text{sgn}(z_1) = -\text{sgn}(z_2)$ necessarily results in a corresponding sine-wave-like long-range lattice distortion. Moreover, taking the switching time as a time of defect evolution from one well-defined sine-wave-like configuration to another, we arrive at a nearly utter absence of "fast" switchings. This makes the deviation from the Arrhenius law all the more pronounced.

## 4. Conclusions

In summary, we made use of molecular dynamics simulation to determine the temperature dependence of the time it takes for the buckled Stone-Wales defect in graphene to switch from one sine-wave-like configuration to another. All switchings observed can be classified as "fast" or "slow", depending on their duration (~ 1 ps or > 10 ps respectively). At $T > 500$ K, the switching time averaged over all (both "fast" and "slow") switching events satisfies the Arrhenius law within the statistical error. At $T = 450$ and 400 K we observed neither "fast" nor "slow" switchings despite that simulation times greatly exceeded the switching times expected from the Arrhenius law. We speculate that this may be a consequence of our assumption of the temperature-independent frequency factor, This point needs further exploration.

## Acknowledgements

We are grateful to M. Maslov for useful discussions. The work was supported in part by RFBR grant No 15-02-02764.

# Literature

1. K.S. Novoselov, A.K. Geim, S.V. Morozov, D. Jiang, Y. Zhang, S.V. Dubonos, I.V. Grigorieva, A.A. Firsov, Science 306 (2004) 666.

2. K.S. Novoselov, A.K. Geim, S.V. Morozov, D.Jiang, M.I. Katsnelson, I.V. Grigorieva, S.V. Dubonos, A.A. Firsov, Nature 438 (2005) 197.

3. Y.Zhang, Y.-W. Tan, H.L. Stormer, P. Kim, Nature 438 (2005) 201.

4. A.H. Castro Neto, F. Guinea, N.M.R. Peres, K.S. Novoselov, A.K. Geim, Rev. Mod. Phys. 81 (2009) 109.

5. A.J. Stone, D.J. Wales, Chem. Phys. Lett. 128 (1986) 501.

6. J.C. Meyer, C. Kisielowski, R. Erni, M.D. Rossell, M.F. Crommie, A. Zettl, Nano Lett. 8 (2008) 3582.

7. J. Kang, J. Bang, B. Ryu, K.J. Chang, Phys. Rev. B 77 (2008) 115453.

8. D.W. Boukhvalov, M.I. Katsnelson, Nano Lett. 8 (2008) 4373.

9. E. Kaxiras, K.C. Pandey, Phys. Rev. Lett. 61 (1988) 2693.

10. L. Li, S. Reich, J. Robertson, Phys. Rev. B 72 (2005) 184109.

11. J. Ma, D. Alfè, A. Michaelides, E. Wang, Phys. Rev. B 80 (2009) 033407.

12. J.C. Meyer, A.K. Geim, M.I. Katsnelson, K.S. Novoselov, T.J. Booth, S. Roth, Nature 446 (2007) 60.

13. M.M. Maslov, A.I. Podlivaev, L.A. Openov, Phys. Lett. A 373 (2009) 1653.

14. L.A. Openov, M.M. Maslov, A.I. Podlivaev, Phys. Lett. A 376 (2012) 3146.

15. X.-L. Sheng, H.-J. Cui, F. Ye, Q.-B. Yan, Q.-R. Zheng, G. Su, J. Appl. Phys. 112 (2012) 074315.

16. L.A. Openov, A.I. Podlivaev, JETP Lett. 90 (2009) 459.

17. L.A. Openov, A.I. Podlivaev, Tech. Phys. Lett. 36 (2010) 31.

18. A.I. Podlivaev, L.A. Openov, Physica E 44 (2012) 894.

19. V.F. Elesin, A.I. Podlivaev, L.A. Openov, Phys. Low-Dim. Struct. 11/12 (2000) 91.

20. A.I. Podlivaev, L.A. Openov, Phys. Solid State 48 (2006) 2226.

21. G.V. Vineyard, J. Phys. Chem. Solids. 3 (1957) 121.

22. M.M. Maslov, L.A. Openov, A.I. Podlivaev, Phys. Solid State 56 (2014) 1239.

23. A.I. Podlivaev, L.A. Openov, Phys. Solid State 57 (2015) 820.